\pdfoutput=1

\documentclass[review,3p,times]{elsarticle}

\usepackage{bm}
\usepackage{amsmath}
\usepackage{mathtools}
\usepackage{dirtytalk}
\usepackage{derivative}
\usepackage{threeparttable}
\usepackage{lscape}
\usepackage{float}
\usepackage[pagewise]{lineno}
\usepackage{booktabs}
\usepackage{threeparttable}
\usepackage{caption}
\usepackage{pgfplots}
\usepackage{subcaption}
\usepackage[short,nocomma]{optidef}
\usepackage{hyperref}
\usepackage{tikz}

\makeatletter
\def\ps@pprintTitle{%
  \let\@oddhead\@empty
  \let\@evenhead\@empty
  \def\@oddfoot{\reset@font\hfil\thepage\hfil}
  \let\@evenfoot\@oddfoot
}
\makeatother

\pgfplotsset{compat=1.18}
\newcommand{\tran}{{\mkern-1.5mu\mathsf{T}}}
\begin{document}

\begin{frontmatter}

\author[1,5]{Sebastián Espinel-Ríos\corref{cor1}}
\ead{sebastian.espinelrios@csiro.au}

\author[1]{José Montaño López}

\author[1,2,3,4]{José L. Avalos}

\cortext[cor1]{Corresponding author}

\affiliation[1]{organization={Department of Chemical and Biological Engineering, Princeton University},
            city={Princeton},
            postcode={08544}, 
            state={New Jersey},
            country={United States}}

\affiliation[2]{organization={Omenn-Darling Bioengineering Institute, Princeton University},
            city={Princeton},
            postcode={08544}, 
            state={New Jersey},
            country={United States}}

\affiliation[3]{organization={The Andlinger Center for Energy and the Environment, Princeton University},
            city={Princeton},
            postcode={08544}, 
            state={New Jersey},
            country={United States}}

\affiliation[4]{organization={High Meadows Environmental Institute, Princeton University},
            city={Princeton},
            postcode={08544}, 
            state={New Jersey},
            country={United States}}
            
\affiliation[5]{organization={Present address: Commonwealth Scientific and Industrial Research Organisation},
            city={Clayton},
            postcode={3168}, 
            state={Victoria},
            country={Australia}}
            
\title{Omics-driven hybrid dynamic modeling of bioprocesses with uncertainty estimation}

\begin{abstract} 
{This work presents an omics-driven modeling pipeline that integrates machine-learning tools to facilitate the dynamic modeling of multiscale biological systems. Random forests and permutation feature importance are proposed to mine omics datasets, guiding feature selection and dimensionality reduction for dynamic modeling. Continuous and differentiable machine-learning functions can be trained to link the reduced omics feature set to key components of the dynamic model, resulting in a hybrid model. As proof of concept, we apply this framework to a high-dimensional proteomics dataset of \textit{Saccharomyces cerevisiae}. After identifying key intracellular proteins that correlate with cell growth, targeted dynamic experiments are designed, and key model parameters are captured as functions of the selected proteins using Gaussian processes. This approach captures the dynamic behavior of yeast strains under varying proteome profiles while estimating the uncertainty in the hybrid model's predictions. The outlined modeling framework is adaptable to other scenarios, such as integrating additional layers of omics data for more advanced multiscale biological systems, or employing alternative machine-learning methods to handle larger datasets. Overall, this study outlines a strategy for leveraging omics data to inform multiscale dynamic modeling in systems biology and bioprocess engineering.}
\end{abstract}

\begin{graphicalabstract}
\centering
\includegraphics[scale=0.52]{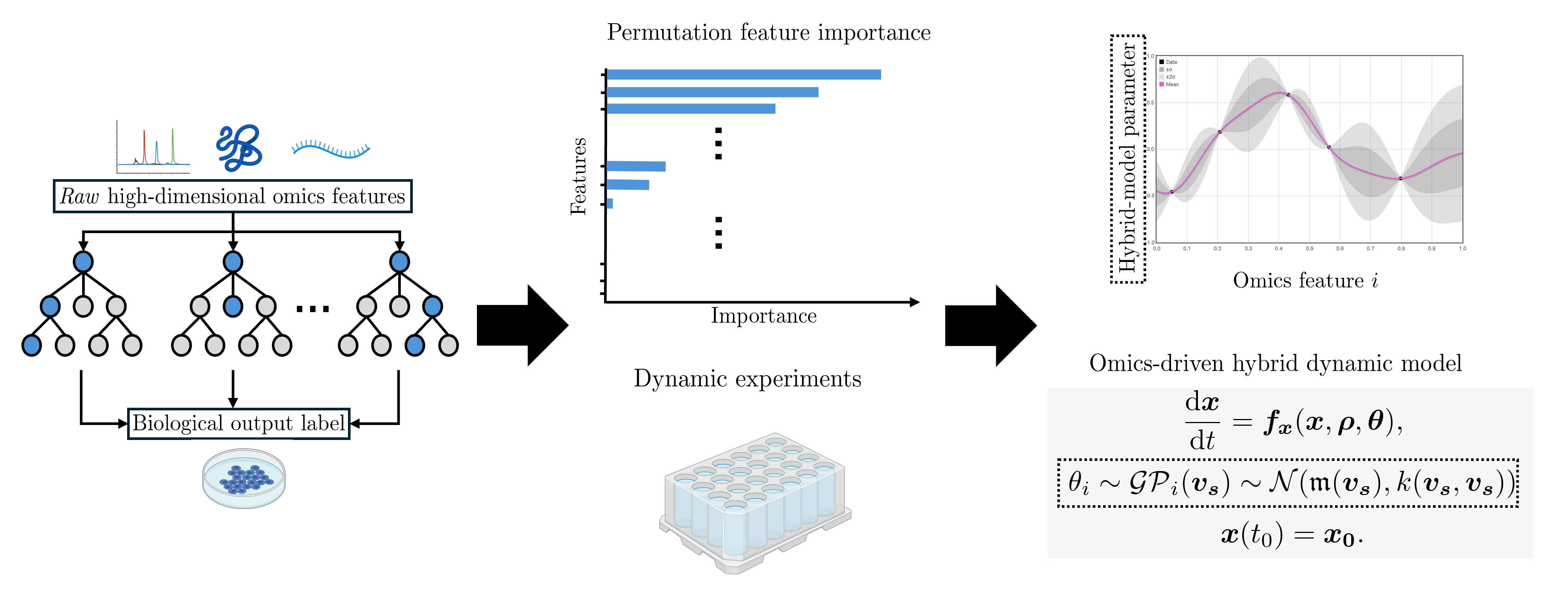}
\end{graphicalabstract}

\begin{highlights} 
\item Mining omics data facilitates dynamic modeling of multiscale biological systems. 
\item Random forests and permutation importance guide feature dimensionality reduction. \item Gaussian-process-predicted model parameters are linked to selected omics features. 
\item Uncertainty estimation leverages the distribution of parameter functions. 
\item A hybrid dynamic model links proteome changes to yeast growth dynamics. \end{highlights}


\end{frontmatter}

\section{Introduction}
\label{sec:introduction}
Systems biology offers methods for creating holistic representations of cells by integrating (high-throughput) experimental data into mathematical/computational models \cite{tavassoly_systems_2018}. However, modeling the dynamics of cells is not a trivial task given their complexity and multiscale nature. Furthermore, the cell phenotype is the outcome of several complex intertwined processes such as signaling, regulation, transcription, translation, and metabolic reactions. In the context of this work, we regard \textit{phenotype} as the set of growth and reaction rates that define the dynamic behavior of the cell in its environment. In biotechnology, the dynamic behavior of a cell can be linked to its production efficiency. Therefore, mathematical modeling is also of great interest to biomanufacturing. Significant efforts are underway to transform the biotechnology industry from empirical and trial-and-error approaches toward smart biomanufacturing, including \textit{model-based} strategies and digitalization \cite{gargalo_towards_2020}. This is in line with paradigms such as quality by design, quality by control, and digital twins \cite{smiatek_towards_2020, tsopanoglou_moving_2021, udugama_role_2023}. Given a specific \textit{quality target product profile}, mathematical models can link (intra)cellular processes to \textit{critical quality attributes} influenced by \textit{critical process parameters}. This can facilitate the understanding, monitoring, optimization, and control of bioprocesses.

\begin{figure*}[htb!]
\centering
\includegraphics[scale=0.4]{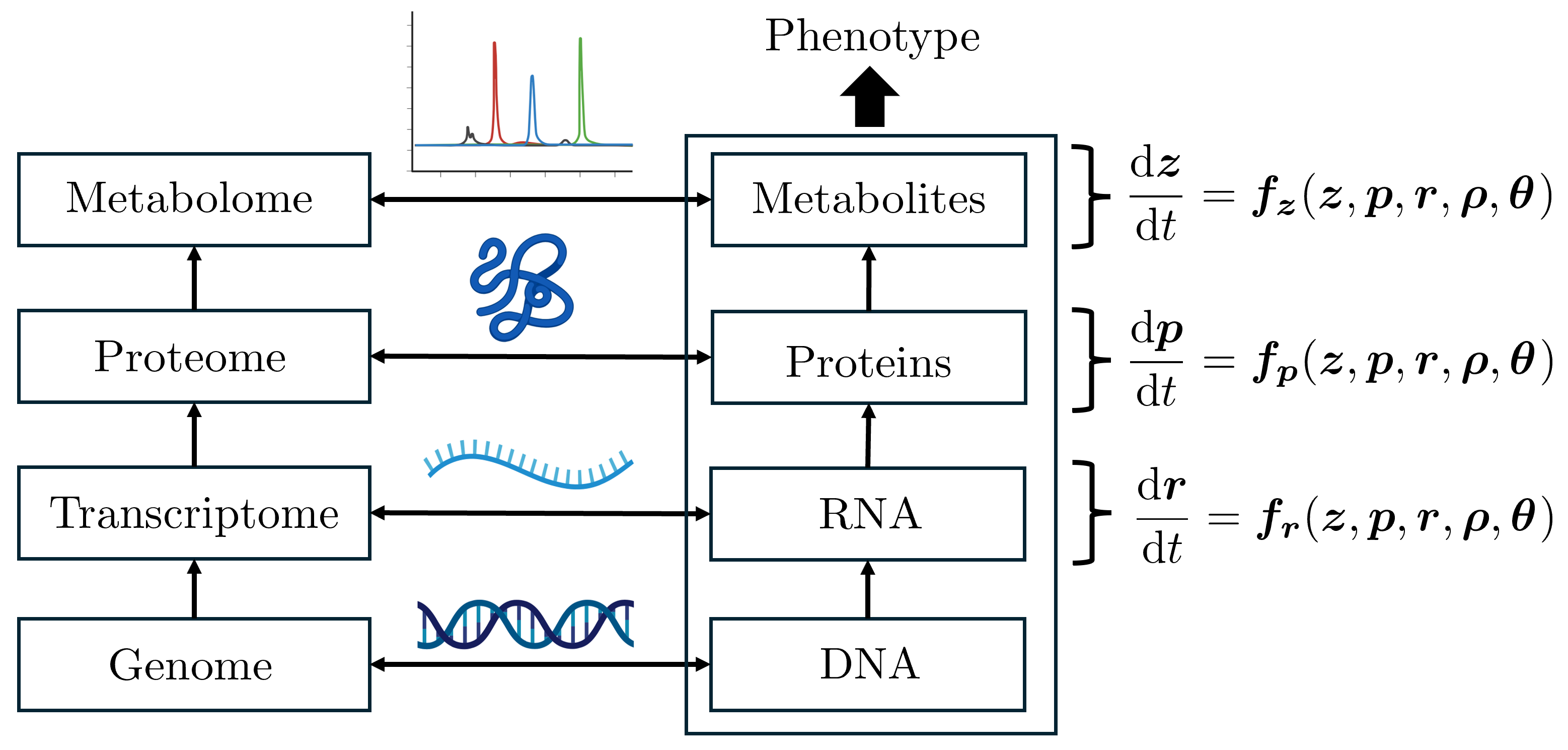}
\caption{Scheme of multi-omics data and its relation to different levels of cellular processes that ultimately determine the cell phenotype. Differential equations can be used to model the complex interactions between these cellular processes, including regulatory mechanisms. For simplicity, the genetic information (genome) is assumed to be \textit{constant} in the cell, hence no dynamic equation is formulated. Refer to Section \ref{sec:materials_methods} for details on notation. The figure contains images from \href{https://www.biorender.com}{https://www.biorender.com}.}
\label{fig:overview}
\end{figure*}

As shown in Fig. \ref{fig:overview}, the sources of experimental data for modeling biological systems can comprise \textit{transcriptomics}, \textit{proteomics}, and \textit{metabolomics} \cite{amer_omics-driven_2021}. The recent availability of automated high-throughput workflows to perform \textit{omics}-based experiments \cite{donati_automated_2023} has the potential to streamline the generation of large datasets for modeling in systems biology. Omics datasets are classified as \textit{big data} since they can be large and high in dimension. In yeast, for example, multi-omics experiments can involve measuring more than 5,000 messenger RNA (mRNA) transcripts and 3,000 proteins \cite{yu_big_2019}. One challenge is how to make sense or \textit{mine} knowledge from omics datasets to create coherent mathematical links between different levels of cellular processes and the resulting phenotype. Being able to predict the \textit{dynamic} cell behavior as a result of changes in the intracellular domain is key to unlocking potential degrees of freedom for improving bioprocess efficiency \cite{espinel-rios_linking_2023,espinel-rios_hybrid_2024}. In addition, \textit{uncertainty} estimation of the model predictions could inform suitable (stochastic) process optimization and control strategies \cite{petsagkourakis_reinforcement_2020,bradford_stochastic_2020}.

In this work, we outline an \textit{omics-driven} pipeline that integrates mechanistic and domain knowledge with machine learning to create hybrid dynamic models (cf. Figure \ref{fig:overview_pipeline}, more details in Section \ref{sec:materials_methods}). In the first step, random forests \cite{breiman2001random} are proposed to reduce the dimensionality of the original feature space into a selected set of features with the highest predictive power. Throughout this work, (omics) \textit{features} refer to \textit{variables}, such as protein concentrations from proteomic analyses, that can be used as \textit{inputs} for building mathematical models. Ranking feature importance is a useful capability of random forests and has been used to highlight key features correlated with specific biological outcomes in the context of biotechnology and systems biology (cf. e.g., \cite{tafur_rangel_silico_2021,messner_proteomic_2023}). In our proposed framework, the labels or outputs of the random forests represent components of the right-hand side of dynamic equations that are difficult to model using purely mechanistic assumptions.

In the second step, we train \textit{continuous} and \textit{differentiable} functions that map the reduced feature dimension previously derived via random forests, and possibly other process parameters, to relevant model components or labels. When embedded in a dynamic model, these data-driven functions allow us to bridge multiple cellular processes within the cell in a straightforward manner.  Direct embedding of random forests into the dynamic model is avoided, as they consist of a series of non-continuous binary decisions, which pose differentiability challenges. Instead, we propose using regressors such as (deep) neural networks for large datasets \cite{Goodfellow-et-al-2016} or Gaussian processes for small to medium datasets \cite{rasmussen_gaussian_2006}. Note that reducing the dimensionality of the features in the first step, using random forests, is crucial as it helps to circumvent the \textit{curse of dimensionality} \cite{altman_curses_2018} when training the differentiable regressors. {Overall, this approach simplifies the construction of multiscale dynamic models by providing a strategic baseline for feature selection and dimensionality reduction when dealing with omics data.}

Two possibilities arise regarding the datasets employed in the modeling pipeline. One approach is to use the same dataset for both feature dimension reduction via random forests and for training the continuous and differentiable functions to be embedded into the hybrid dynamic model. This option assumes that the experimental and analytical conditions used to generate the dataset are compatible with the desired application setup. Alternatively, as in the case study of this work, one can leverage available multi-omics datasets to identify the most important omics features correlated with the biological outputs of interest via random forests (\textit{first step}), even if the conditions for generating the omics dataset differ from the desired application setup. With this information, new experiments can then be designed to explore varying values of the selected features, allowing for the training of specific continuous and differentiable model components for integration into the hybrid dynamic model (\textit{second step}). This approach of mining available data for feature selection can significantly save time and resources, enabling the reuse of \textit{knowledge} captured by previous omics experiments.

The aim of this work is to outline the core steps of the proposed omics-driven hybrid dynamic modeling pipeline. As proof of concept, we use a proteomics dataset derived from the single-gene knockout collection of \textit{Saccharomyces cerevisiae}, along with the reported normalized growth rate of each deletion strain \cite{messner_proteomic_2023}. Using random forests, we identify the most important proteins whose intracellular concentrations correlate with cell growth. {Upon conducting growth kinetic experiments,} we train Gaussian processes that predict key model parameters of the dynamic model as \textit{functions} of the selected intracellular protein concentrations. {This bridges the feature selection and dimensionality reduction step via random forests to the construction of the Gaussian-process-supported multiscale dynamic model}. Gaussian processes provide both the mean and variance of the predictions, and we leverage this variance to estimate the uncertainty in the dynamic model’s predictions. {In this work, we selected Gaussian processes as the continuous and differentiable machine-learning functions to implement in the hybrid model driven by the small experimental dataset of kinetic experiments available.}

\section{Methodology}
\label{sec:materials_methods}
\subsection{Overview of the case study and modeling pipeline}
\label{subsec:overview_methodology}
\begin{figure*}[htb!]
  \includegraphics[scale=0.5]{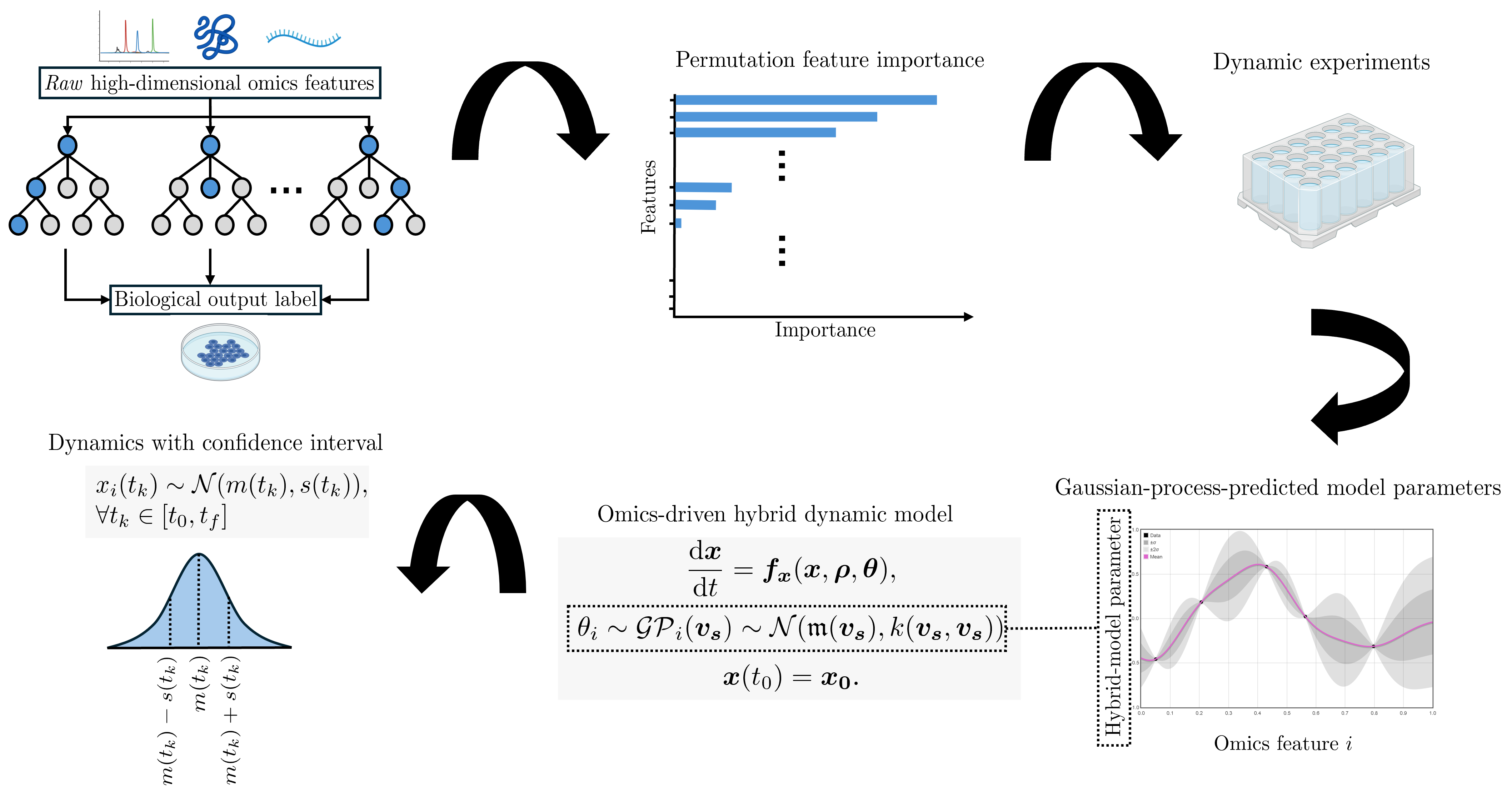}
   \caption{Pipeline of the proposed omics-driven hybrid modeling framework. {The black arrows indicate the general flow of the pipeline}. Random forests are used to rank feature importance from \textit{raw} high-dimensional omics data, resulting in a set of \textit{selected features}. Experiments can be designed to explore the effects of changes in these selected features on dynamic cell behavior. Gaussian processes are then employed to link key parameters of the dynamic model to changes in the selected features, resulting in a hybrid model. The uncertainty from the Gaussian processes is propagated into the time domain, enabling the estimation of uncertainty in the predicted dynamic trajectories. Refer to Section \ref{sec:materials_methods} for details on notation. The figure contains images from \href{https://www.biorender.com}{https://www.biorender.com}. The image of the Gaussian process was generated using the demo tool available at \href{http://chifeng.scripts.mit.edu/stuff/gp-demo/}{http://chifeng.scripts.mit.edu/stuff/gp-demo/}.}
  \label{fig:overview_pipeline}
\end{figure*}

For clarity of presentation, let us provide an overview of the case study and methodology (Fig. \ref{fig:overview_pipeline}), which will be described in more detail in the following sections.

The proteome dataset \cite{messner_proteomic_2023} used in this study, captures the effect of non-essential single-gene deletions in \textit{S. cerevisiae} {on growth rates normalized with respect to the wild-type (WT) strain}. In \cite{messner_proteomic_2023}, {\textit{normal growth} is defined in the range of 0.9-1.0, while \textit{slow growth} is defined in the range of 0.3-0.4}. {Thus, \textit{moderate} growth would be in between the latter ranges. Note that some growth rates can be higher than 1 if, for example, the strain grows faster than the WT}. The growth rates were determined from growth curves based on pixel intensity upon imaging plates with agar media (more details in \cite{messner_proteomic_2023}). In that study, random forests were applied to predict growth rates from the proteome dataset and to estimate feature importance. Similarly, in our work, random forests are used to estimate feature importance, but we follow a slightly different and more structured approach (see below). Additionally, in our approach, the generation of random forests serves as an intermediate step toward the end goal of building a hybrid dynamic model.

Rather than using the entire dataset, we randomly sample the proteome data to ensure equal representation across the full range of normalized growth rates. To achieve this, we split the original data into two sets based on the median growth rate (in synthetic minimal medium) and sample 350 proteome profiles, each corresponding to a different single-gene deletion. Note that each proteome sample contains the concentration of 1,850 intracellular proteins. For repeated proteome profiles of a given knockout, the mean proteome is computed. This random yet representative sampling approach addresses the skewness in the original dataset \cite{messner_proteomic_2023}, which was dominated by higher growth rates, thus reducing potential bias during training and making the computational load more manageable. We denote the omics dataset of 350 proteome profiles as $\mathcal{D} \in \mathbb{R}^{n_d \times (n_v + 1)}$, where $n_d$ is the total number of samples, $n_v$ is the number of features (total measured omics state variables, i.e., protein concentrations), and 1 represents the relevant biological output label of interest (normalized growth rate). The sampled dataset $\mathcal{D}$ is then split into training and test subsets, with 80 \% of the data used to train the random forest and 20 \% reserved to evaluate the model's generalization performance on unseen data. Another key difference between our approach and the original study \cite{messner_proteomic_2023} is that, while the original study primarily used default hyperparameters for training random forests, we perform a grid search to optimize them. Additionally, whereas the original study estimated feature importance based on the increase in (Gini) node purity for each protein from the \textit{training} subset, we use feature permutation importance. The latter assesses the impact of shuffling each protein’s data in the \textit{test} subset on the model accuracy, making it independent of the training process and providing a more reliable measure of feature importance.

We then assess the effect of increasing the number of features (e.g., protein concentrations in proteomics), ranked by importance, on the random-forest model’s prediction accuracy, evaluated using the test subset. From this analysis, we select the minimum set of features with a model's coefficient of determination ($R^2$) that is at least equal to the accuracy of the random forest using the full feature dimension from the proteome dataset. Throughout this work, we refer to this set of features as the \textit{selected (most important) features}. 

It is worth noting that the original proteome dataset \cite{messner_proteomic_2023} was based on experiments performed on agar (\textit{solid}) media, whereas biotechnological production processes are typically conducted in liquid media. Additionally, the available data does not include \textit{dynamic} concentrations. Therefore, to exemplify the construction of the hybrid dynamic model, we conducted seven kinetic experiments in liquid media: six with strains that have gene deletions from the test subset, and one with the WT strain as a reference. These knockout strains have predicted growth rates (on agar medium) that are well-distributed across the dataset's range.

With this assumption, we fit dynamic models for each of the strains mentioned above and link the resulting parameter values to the selected features using Gaussian processes. This is where the feature selection process connects to the hybrid dynamic modeling. The structure of the proposed dynamic model bridges the proteome layer with the macro-scale growth behavior. The underlying assumption is that the concentrations of these selected features sufficiently correlate with the parameters of the dynamic growth model. Finally, we sample parameter values from the distributions predicted by the Gaussian processes to generate dynamic state trajectories, enabling the estimation of confidence intervals for the hybrid dynamic model.

We acknowledge that, for a more robust and comprehensive model, additional experiments would be necessary to explore a broader range of the selected feature space and to confidently predict beyond the training set of Gaussian-process-predicted parameters. However, it is important to emphasize that the primary aim of this work is to demonstrate the \textit{methodology} of the omics-driven hybrid dynamic modeling pipeline, with the case study serving as an illustrative example. Naturally, the outlined methodology can be expanded or adapted to other scenarios or biological systems involving more or different omics data. As is typical with machine learning approaches, the methodology would also benefit from larger experimental or process datasets, enhancing its predictive power and generalizability.

\subsection{General modeling structure}
\label{sec:framework}
In a general form, we represent the dynamics of a cell through the following system of differential equations ({cf. Fig. \ref{fig:overview})}:
\begin{equation}
\frac{\mathrm{d}\bm{z}}{\mathrm{d}t} = \bm{f_z}(\bm{z}, \bm{p}, \bm{r}, \bm{\rho}, \bm{{\theta_c}}, \bm{{\theta_f}}),
\label{eq:ode_z}
\end{equation}
\begin{equation}
\frac{\mathrm{d}\bm{p}}{\mathrm{d}t} = \bm{f_p}(\bm{z}, \bm{p}, \bm{r}, \bm{\rho}, \bm{{\theta_c}}, \bm{{\theta_f}}),
\label{eq:ode_p}
\end{equation}
\begin{equation}
\frac{\mathrm{d}\bm{r}}{\mathrm{d}t} = \bm{f_r}(\bm{z}, \bm{p}, \bm{r}, \bm{\rho}, \bm{{\theta_c}}, \bm{{\theta_f}}),
\label{eq:ode_r}
\end{equation}
\begin{equation}
\bm{x} := \left[\bm{z}^\tran, \bm{p}^\tran, \bm{r}^\tran\right]^\tran,
\label{eq:x_def}
\end{equation}
\begin{equation}
\bm{x}(t_0) = \bm{x_0},
\label{eq:x0}
\end{equation}
where $\bm{z} \in \mathbb{R}^{n_z}$, $\bm{p} \in \mathbb{R}^{n_p}$, and $\bm{r} \in \mathbb{R}^{n_r}$ represent concentrations of metabolites, proteins, and mRNA transcripts, respectively. Cell biomass, as a single lumped component, can be also considered in $\bm{z}$. Here, $t$ is the process time, and $t_0$ is the time at the initial condition. Throughout this work, we omit the time-dependency of variables, unless unclear from the context.  For simplicity of notation, $\bm{x} \in \mathbb{R}^{n_z+n_p+n_r}$ comprises all the dynamic states. 

Furthermore, the functions $\bm{f_i}: \mathbb{R}^{n_z} \times \mathbb{R}^{n_p} \times \mathbb{R}^{n_r} \times \mathbb{R}^{n_\rho} \times \mathbb{R}^{n_{\theta_c}} \times \mathbb{R}^{n_{\theta_f}} \rightarrow \mathbb{R}^{n_i}, \, \forall i = \{\bm{z}, \bm{p}, \bm{r}\}$ describe the right-hand side of the dynamic equations. This can include both a knowledge-based/mechanistic part and a machine-learning/data-driven part, resulting in hybrid functions in this case. Process variables, such as environmental conditions, disturbances, or control inputs, are captured in $\bm{\rho} \in \mathbb{R}^{n_\rho}$. In the general case, constant model parameters are represented by $\bm{{\theta_c}} \in \mathbb{R}^{n_{\theta_c}}$, while other parameters $\bm{{\theta_f}} \in \mathbb{R}^{n_{\theta_f}}$ are considered as \textit{functions} of other dynamic states and process variables, $\bm{{\theta_f}}: \mathbb{R}^{n_z} \times \mathbb{R}^{n_p} \times \mathbb{R}^{n_r} \times \mathbb{R}^{n_\rho} \times \mathbb{R}^{n_{\theta_c}} \rightarrow \mathbb{R}^{n_{\theta_f}}$, and thus are allowed to vary. For simplicity of notation, $\bm{\theta}:=[\bm{{\theta_c}}^\tran, \bm{{\theta_f}}^\tran]^\tran$. In the case study, we assume that the parameters in $\bm{{\theta_f}}$ are functions of selected features $\bm{v_s}\in \mathbb{R}^{n_{v_s}}$, in particular, selected intracellular protein concentrations. Specifically, the parameter functions are modeled using Gaussian processes ${\theta_f}_i \sim \mathcal{GP}_i(\bm{v_s}),\, \forall i \in \{1,2...,n_{\theta_f}\}$ (more details in Section \ref{subsec:framework_gps}).

Without loss of generality, we assume that the genome encodes \textit{constant} or \textit{static} information, thus this omics layer is not modeled. If integrating epigenomics data (changes in the genome due to environmental conditions) is desired, another dynamic equation can be formulated. Including a genome model layer can be particularly relevant for biotechnological systems like Chinese Hamster Ovary (CHO) cells. In these systems, specific culture conditions have been observed to influence genome sequence and DNA methylation \cite{feichtinger_comprehensive_2016}.

\subsection{Dimension reduction of omics features using random forests}
\label{subsec:framework_forest}
A random-forest model is a collection of decision trees that can capture non-linear relationships between features and labels. In this section, we provide an overview of random forests for regression tasks. For more detailed information, the reader can refer to \cite{breiman2001random,cutler2012,aldrich_process_2020}. The output label of random forests is computed as the average output value of the decision trees:
\begin{equation}
\hat{l}(\bm{v}) = \frac{1}{n_T} \sum_{i=1}^{n_T} T_{\mathcal{D}_i} (\bm{v}),
\end{equation}
where $T_{\mathcal{D}_i} (\bm{v})$ is a decision tree $i$ out of $n_T$ decision trees. The feature and predicted label vectors are denoted as $\bm{v} \in \mathbb{R}^{n_v}$ and $\hat{l} \in \mathbb{R}$, respectively. Note that $\bm{v}$ in the case study represents the full proteomics features, i.e., protein concentrations, from which we aim to extract the selected features $\bm{v_s}$. In a decision tree, a given data entry from the dataset is partitioned in a binary manner, starting from the \textit{root} node of the tree to \textit{child} nodes, following conditional rules based on the feature values. These child nodes can in turn become \textit{parent} nodes for further child nodes. The output label associated with a given leaf node (with no further partitioning) corresponds to the average label value of the training samples that reached that node.

During the training process of a decision tree, the algorithm selects the best constraint for a given feature such that an \textit{impurity} criterion is minimized at each split. In this work, we use the weighted mean square error (MSE) as a criterion:
\begin{equation}
\text{MSE}_\mathrm{children}^{(s)} = \frac{n_L}{n_L+n_R} \cdot \text{MSE}_L^{(s)} + \frac{n_R}{n_L+n_R} \cdot \text{MSE}_R^{(s)},
\label{eq:criterion}
\end{equation}
\begin{equation}
\text{MSE}_i^{(s)} = \frac{1}{n_i} \sum_{j=1}^{n_i} (l_j - \hat{l}_j)^2, \, \forall i \in \{L, R\}.
\end{equation}
Here, the indices $L$ and $R$ indicate the left and right child nodes, respectively. The superscript $(\cdot)^{(s)}$ indicates the split event $s$. The number of samples in the respective left and right child nodes is denoted by $n_L$ and $n_R$. Furthermore, $l_j$ and $\hat{l}_j$ represent the actual and predicted label values, respectively, in the corresponding child node.

We can influence the performance of a random forest by selecting proper hyperparameters. In the case study, when applicable, hyperparameters were optimized using a grid-search approach. Specifically, we considered 100, 300, and 500 decision trees, maximum depths of 20, 30, and 50, a minimum number of samples required to split a node of 2, 5, 7, and 10, and a minimum number of samples per leaf node of 1, 2, 3, and 5. The number of features considered for splitting a node was computed using both $\log_2(n_v)$ and $\sqrt{n_v}$. Additionally, we tested training with and without bootstrapping (training data sampled with replacement from the original training set). Using 5-fold cross-validation, the best hyperparameter set was chosen based on the mean validation scores across the folds. Note that this validation score is calculated from the validation data, split from the training subset, in the cross-validation folds.

To rank the key features determining the regression label, we use the concept of permutation importance. It measures how much the $R^2$ decreases when a feature is permuted or randomly shuffled on the \textit{test} dataset. The permutation importance of a feature $v_j$, $\mathrm{PI}{v_j}$, is calculated as: \begin{equation} 
\mathrm{PI}{v_j} = R_u^2 - \frac{1}{n_S} \sum_{p=1}^{n_S} R_{s_{j,p}}^2, \, \forall j \in \{1, 2, \dots, n_v\}, 
\end{equation} where $R_u^2$ is the model's $R^2$ score on the unshuffled test data, $R_{s_{j,p}}^2$ is the $R^2$ score after permuting feature $v_j$ during the $p$-th permutation, and $n_S$ is the total number of permutations or shufflings.  By evaluating the impact of each feature's shuffling on the $R^2$ score, we can rank the features by their importance. Features causing larger decreases in $R^2$ when shuffled are considered more important. {This is often correlated with how much the decision trees rely on that feature when making predictions. If the feature is useful for correlating to the regression output, e.g., if it helps to minimize the impurity criterion, the decision trees are likely to use these features during the splits. However, note that permutation importance reflects the \textit{unique} contribution of a feature. Shuffling two or more features simultaneously may lead to a larger drop in $R^2$ due to synergistic effects between features. Nevertheless, the ranking derived from permutation importance offers a practical approach for estimating the relevance of features, in particular in high-dimensional spaces.} After a suitable analysis (see Section \ref{subsec:feature_ranking}), a set of features $\bm{v_s}$ can be selected and used {for} building the hybrid dynamic model.

Although in this work we use random forests for dimension reduction of the feature space, it is worth noting that other options are also possible. One alternative is the use of principal component analysis (PCA) \cite{jolliffe_principal_2002} as in \cite{zelezniak_machine_2018}. Although PCA can be fast and simple to implement, it has several disadvantages when compared to random forests. First, the dimension reduction is done without considering the output label. That is, the principal components \textit{explain the variance in the feature space} but do not explicitly link to their predictive power. Furthermore, PCA assumes that the principal components are \textit{linear} combinations of the original features, which may fail to capture non-linear patterns in the data.

\subsubsection{Fermentation experiments}
Dynamic experiments in liquid media for modeling the growth of \textit{S. cerevisiae}, under varying intracellular protein concentrations (selected features), were performed as follows. A colony from selected single-gene deletion strains was grown overnight in 10 mL of synthetic complete medium (SC) supplemented with 2 \% glucose in a 50 mL conical tube (Cat. 229421, CELLTREAT Scientific Products, MA, USA). The cultures were incubated in an orbital shaker (Eppendorf, New Brunswick, USA) at 30 $^{\circ}$C and 200 rpm. The next day, overnight cultures were used to inoculate 1 mL ({five replicates}) of SC + 2 \% glucose in 24-well plates (CELLTREAT non-treated sterile flat-bottom plates). The cultures were incubated at 30 $^{\circ}$C in a TECAN plate reader (Infinite M200PRO), which performed optical density ($\mathrm{OD_{600}}$) measurements every 30 minutes. Upon serial dilutions, a calibration curve was constructed to fit the TECAN readings to \textit{real} $\mathrm{OD_{600}}$ values, correcting for deviations at high optical densities. Throughout this work, we report the corrected values. 

\subsection{Gaussian-process-predicted parameters with reduced features}
\label{subsec:framework_gps}
For each dynamic experiment, a specific model structure is proposed and the corresponding \textit{nominal} model parameters are estimated (see Section \ref{subsec:model_structure_r}). Gaussian processes are used to map the parameter values to the selected omics features, i.e., the selected intracellular protein concentrations $\bm{v_s}$. For each parameter function ${\theta_f}_i,\, \forall i \in \{1,2...,n_{\theta_f}\}$, we denote with $\bm{L} \in \mathbb{R}^{n_{e}}$ the training dataset for the output label of the Gaussian process. Similarly, we denote with $\bm{V} \in \mathbb{R}^{(n_{v_s} \times n_{e})}$ the training dataset for the input features of the Gaussian process. The total number of dynamic experiments is given by $n_e$. Each parameter function ${\theta_f}_i(\bm{v_s}) \sim \mathcal{GP}_i(\bm{v_s})$ has a \textit{prior} normal distribution with mean function $\mathfrak{m}: \mathbb{R}^{n_{v_s}} \rightarrow \mathbb{R}$ and covariance function $k: \mathbb{R}^{n_{v_s}} \times \mathbb{R}^{n_{v_s}} \rightarrow \mathbb{R}$, i.e., $\mathcal{GP}_i(\bm{v_s}) \sim \mathcal{N}(\mathfrak{m}(\bm{v_s}), k(\bm{v_s},\bm{v_s}))$, before observing any data. {A prior mean of zero is often employed by default if there is no prior \textit{guess} or \textit{belief} about the \textit{true} mean function. In this work, the zero prior mean is used unless otherwise stated}. Here, we will provide a general overview of Gaussian processes for regression tasks. For more detailed information, the reader can refer to \cite{rasmussen_gaussian_2006}.

The Gaussian process under consideration models a \textit{true} function ${\theta_f}_i: \mathbb{R}^{n_{v_s}} \rightarrow \mathbb{R}$ from \textit{noisy} observations $\kappa \in \mathbb{R}$, where the noise is normally distributed $\epsilon \sim \mathcal{N}(0,\sigma_n^2)$:
\begin{equation}
\kappa = {\theta_f}_i(\bm{v_s}) + \epsilon.
\end{equation}

The kernel matrix $\bm{K} \in \mathbb{R}^{n_e \times n_e}$ captures the \textit{neighborhood} of the features:
\begin{equation}
\bm{K}=\begin{bmatrix}
  k(\bm{v_{s_1}},\bm{v_{s_1}})  & \cdots  & k(\bm{v_{s_1}},\bm{v_{s_{n_e}}}) \\ 
\vdots & \ddots & \vdots\\ 
 k(\bm{v_{s_{n_e}}},\bm{v_{s_1}}) & \cdots &  k(\bm{v_{s_{n_e}}},\bm{v_{s_{n_e}}})
\end{bmatrix},
\end{equation}
where the element $(i,j)$ corresponds to the kernel function $k(\bm{v_{s_i}},\bm{v_{s_j}})$ of two inputs $\bm{v_{s_i}} \in \mathbb{R}^{n_{v_s}}$ and $\bm{v_{s_j}} \in \mathbb{R}^{n_{v_s}}$. In this work, we {consider the commonly used} squared exponential kernel function \cite{MOWBRAY2021108054}:
\begin{equation}
    k(\bm{v_{s_i}},\bm{v_{s_j}}|\bm{\tau}) = \sigma^2 \mathrm{exp}\left( \frac{-(\bm{v_{s_i}}-\bm{v_{s_j}})^\tran(\bm{v_{s_i}}-\bm{v_{s_j}})}{2d^2}\right),
\label{eq:kernel_function}
\end{equation}
where $\sigma^2 \in \mathbb{R}$ is the signal variance and $d\in \mathbb{R}^{n_{v_s}}$ is the length-scale, with independent length-scales for the features (automatic relevance determination). These are hyperparameters of the Gaussian process, which are contained in $\bm{\tau}$, along with the noise variance $\sigma_n^2$. All hyperparameters are optimized by maximizing the log marginal likelihood:
\begin{equation}
\log(\mathrm{P}(\bm{L}|\bm{V},\bm{\tau}))=-\frac{1}{2}\bm{L}^\tran(\bm{K}+\sigma_n^2\bm{I_d})^{-1}\bm{L} -\frac{1}{2}\mathrm{log}\left(|\bm{K}+\sigma_n^2\bm{I_d}|\right) - \frac{n_e}{2}\mathrm{log}(2\pi),
\end{equation}
where $\bm{I_d}$ is a general identity matrix of appropriate dimension. {To facilitate convergence, we employed \textit{heuristic} approaches for the \textit{initial guesses} of the hyperparameters, including the sample variance of the label values for the signal variance and 10–30 \% of the range of the feature values for the length scale.  The \textit{initial guess} for the noise variance was set to the average variance derived from the confidence intervals of the parameter estimates (${\theta_f}_i$) and adjusted when necessary to improve convergence.}

The conditional posterior of the Gaussian process, i.e., after \textit{observing} data, given a test input $\bm{v_s^*} \in \mathbb{R}^{n_{v_s}}$, follows a normal distribution $\mathrm{P}(\mathcal{GP}_i(\bm{v_s}^*)|\bm{V},\bm{L}) \sim \mathcal{N}(\bar{\mathfrak{m}},\bar{\mathfrak{s}})$, where $\bar{\mathfrak{m}}$ represents the predictive (posterior) mean and $\bar{\mathfrak{s}}$ denotes the predictive (posterior) variance, defined by:
\begin{equation}
\bar{\mathfrak{m}}(\bm{v_s}^*)=\bm{\Tilde{k}}^\tran(\bm{K}+\sigma_n^2\bm{I_d})^{-1}\bm{L},
\end{equation}
\begin{equation}
\bar{\mathfrak{s}}(\bm{v_s}^*) = k(\bm{v_s}^*,\bm{v_s}^*) - \bm{\Tilde{k}}^\tran(\bm{K}+\sigma_n^2\bm{I_d})^{-1}\bm{\Tilde{k}},
\end{equation}
\begin{equation}
\bm{\Tilde{k}} := [k(\bm{v_{s_1}},\bm{v_s}^*),...,k(\bm{v_{s_{n_e}}},\bm{v_s}^*)]^\tran.
\end{equation}
Thus, the Gaussian process predicts \textit{distributions over functions}, which can be used to estimate uncertainty.

\subsection{Uncertainty estimation of the hybrid model}
\label{subsec:framework_bootstrapping}
We sample each parameter from the normal distribution generated by the trained Gaussian processes and simulate the model dynamics $n_\mathrm{sim}=30$ times. Note that the initial condition of the model is also sampled from a normal distribution with mean and standard deviation computed from the replicate experiments. For each dynamic state, the trajectories are used to estimate the confidence interval of the model’s predictions. At each time step, we calculate the mean and standard deviation across all simulations. With $\alpha = 0.05$, the 95\% confidence level interval is computed by applying the \textit{z}-critical value from the normal distribution. For a dynamic state $x_i$ at time $t_k$, the confidence interval is given by:
\begin{equation}
\mathrm{CI} = m(t_k) \pm z_{\alpha/2} \cdot \frac{s(t_k)}{\sqrt{n_\mathrm{sim}}},
\end{equation}
where $m(t_k)$ and $s(t_k)$ are the mean trajectory and the standard deviation of the trajectories at that time step.

\subsection{Numerical methods}
In the case study, random forests were implemented in Python using scikit-learn \cite{scikit-learn}. Parameter estimation for the dynamic experiments was performed using the particle swarm algorithm in COPASI \cite{hoops_copasicomplex_2006}. The training of the Gaussian processes and the dynamic simulations of the hybrid model were conducted using HILO-MPC \cite{pohlodek_flexible_2024}.

\section{Results and discussion}
\label{sec:case}
\subsection{Feature selection leveraging random forests}
\label{subsec:feature_ranking}
\begin{figure*}[htb!]
    \begin{subfigure}{0.48\textwidth}
      \captionsetup{justification=centering}
      \includegraphics[scale=0.68]{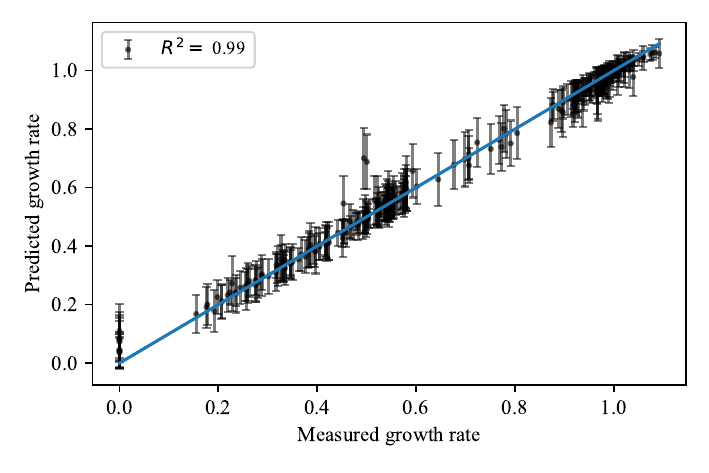}
      \subcaption[]{Parity plot: training set}
      \label{fig:parity_plot_train}
    \end{subfigure}
    \begin{subfigure}{0.48\textwidth}
      \captionsetup{justification=centering}
      \includegraphics[scale=0.68]{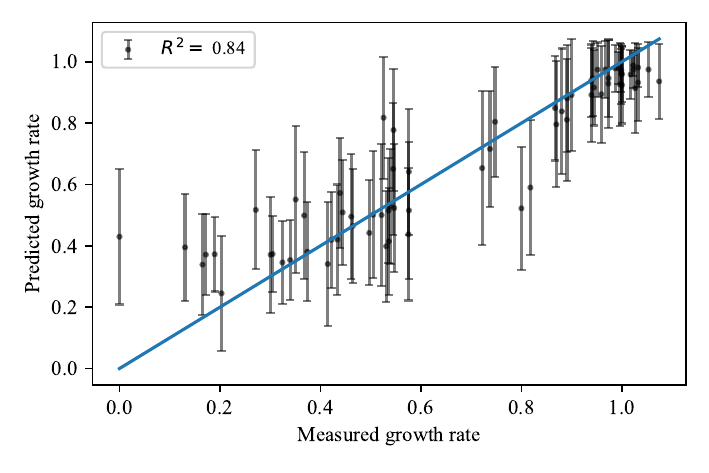}
      \subcaption[]{Parity plot: test set}
      \label{fig:parity_plot_test}
    \end{subfigure}
    \caption{Parity plots evaluated on the A) training and B) test subsets for the random forest model with optimized hyperparameters. Number of trees: 100, number of features considered for splitting at each node: $\sqrt{n_v}$, maximum depth: 20, minimum samples for node splitting: 7, minimum samples for leaf nodes: 2, and bootstrapping: false.  {The prediction uncertainty is represented by the standard deviation of the predictions from the individual decision trees within the random forest.}}
    \label{fig:parity_plots_all}
\end{figure*}

In Figs. \ref{fig:parity_plot_train} and \ref{fig:parity_plot_test}, we present the parity plots of the random forest with optimal hyperparameters using both the sampled training and test datasets. The $R^2$ on the training and test sets are 0.99 and 0.84, respectively. The almost \textit{perfect} fit of the training set was expected as the model was trained on this data. More informative is the fit on the test set (unseen data by the trained model). In this case, a value of 0.84 shows a \textit{strong} generalization capability of the random forest. This value is higher than that achieved in the work of \cite{messner_proteomic_2023}, from which the sampled proteome dataset was obtained. This improvement can be attributed to the fact that our sampled dataset has a more balanced distribution over the entire range of label values, combined with the performed hyperparameter optimization. It is worth noting that the standard deviation of the predictions is larger on the test set than on the training set, despite the \textit{good} \textit{mean} correlation ($R^2 = 0.84$). This outcome was also expected, as models are typically less certain when making predictions on \textit{unseen} data.

\begin{figure*}[htb!]
  \includegraphics[scale=0.82]{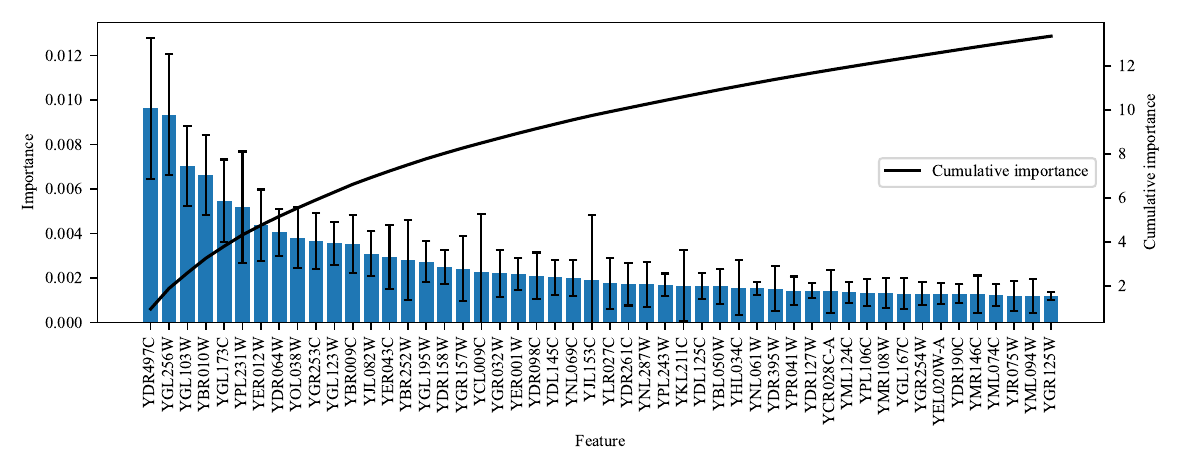}
  \caption{Top 50 important features (intracellular protein concentrations) according to the permutation importance metric computed from the trained random forest with optimal hyperparameters. The cumulative importance is also shown.}
  \label{fig:feature_importance_plot}
\end{figure*}

\begin{figure*}[htb!]
\centering
  \includegraphics[scale=0.68]{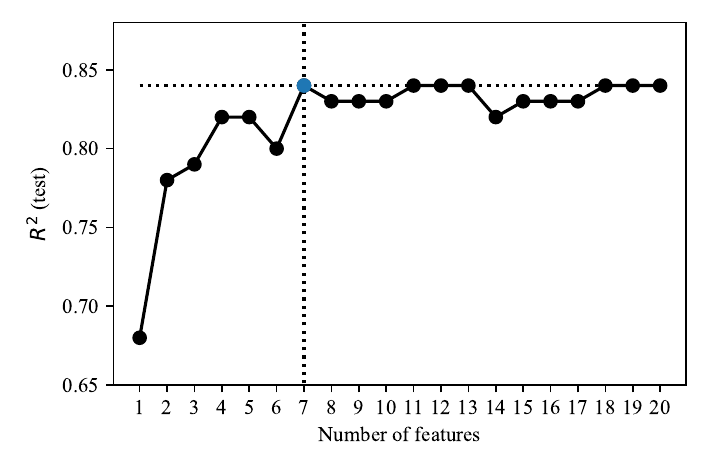}
  \caption{Increasing number of important features (cf. Fig. \ref{fig:feature_importance_plot}), up to 20 features, on the $R^2$ value of the random-forest models evaluated using the test set. Each random forest followed individual grid-search hyperparameter optimization. The intersection (in blue) of the dotted lines denotes the \textit{earliest} $R^2$ value that is equivalent to that of the optimized random forest using all the features.}
  \label{fig:feat_coeff_det_bar_plot}
\end{figure*}

{For clarity of presentation, we show in Fig. \ref{fig:feature_importance_plot}} the top 50 most important protein features, according to the permutation importance metric computed from the trained random forest referred in Fig. \ref{fig:parity_plots_all}. Throughout this study, protein names follow the standard \textit{S. cerevisiae} open reading frame (ORF) nomenclature for the encoding genes. We selected the minimum number of important features, in descending order, that were necessary to achieve (at least) the $R^2$ value of the random forest evaluated on the test dataset using the full feature dimension, i.e., $R^2 = 0.84$. We found that the first seven important features, i.e., YDR497C (\textit{ITR1}), YGL256W (\textit{ADH4}), YGL103W (\textit{RPL28}), YBR010W (\textit{HHT1}), YGL173C (\textit{XRN1}), YPL231W (\textit{FAS2}), and YER012W (\textit{PRE1}), were sufficient for this purpose (Fig. \ref{fig:feat_coeff_det_bar_plot}). These protein concentrations were included in the vector $\bm{v_s}$, hence $n_{v_s} = 7$. This represents only about 0.4 \% of the original feature dimension, which is a significant reduction and facilitates the building of the hybrid dynamic model. This also suggests that, in principle, only these protein concentrations would need to be measured for accurately predicting growth, potentially simplifying process monitoring and reducing the required analytical measurements.

{The improvement in the performance of the random forest with an increasing number of features is driven by synergistic interactions between them. The analysis in Fig. \ref{fig:feat_coeff_det_bar_plot} focuses on the $R^2$ value evaluated on the test set. The latter remained unseen by the model during training. The test set was also not used in the cross-validation (hyperparameter optimization), which was conducted on a subset of the training data, as mentioned in Section \ref{subsec:framework_forest}. Therefore, the test set better reflects the model's predictive performance compared to using the training set.} 

One important consideration is whether the concentrations {of the proteins in $\bm{v_s}$} are merely proxies for the growth rate (correlation) or if they actually determine growth (causation). It is essential to recognize that the results provided by the random forest models demonstrate, \textit{at most}, a correlation. However, this does not diminish the benefits of the approach. Even if we assume only correlation, linking these intracellular protein concentrations to growth dynamics within multiscale modeling frameworks would still be highly valuable. Such an approach integrates cellular processes in a more holistic manner, which is a core goal in systems biology. For example, correlating key intracellular protein concentrations could bridge the proteome layer with observable bioprocess-level behavior.

In cases where these features are \textit{manipulatable} variables for advanced applications (once causation is proven), this dimension reduction approach becomes critical in guiding the efficient design and engineering of biological systems, such as in dynamic metabolic engineering. In the case study, proving causation would require tailored experiments, such as modulating selected protein concentrations and evaluating their effects on growth rate. The exploration space would already be reduced, as only the most important features would be considered for further analysis. If causation is demonstrated (which is beyond the scope of this work), it would be worth exploring the potential of manipulating a subset of the selected proteins to modulate cell growth in yeast. For instance, one could design two-stage processes where the concentration of these proteins maximizes cell growth in the first phase, followed by a production phase where the concentration of these proteins is adjusted to reduce or halt growth.

We used the \textit{Saccharomyces} Genome Database (SGD) \cite{cherry_saccharomyces_2012} to investigate the function of the selected features. YDR497C (\textit{ITR1}) is a transmembrane myo-inositol transporter. Myo-inositol is incorporated into phosphoinositides and inositol phosphates, which are involved in signaling, regulatory, and structural functions \cite{suliman_myo-inositol_2021}. YGL256W (\textit{ADH4}) is an alcohol dehydrogenase that catalyzes the reduction of acetaldehyde to ethanol during the fermentation of glucose, thus participating in the central carbon metabolism. YGL103W (\textit{RPL28}) is a ribosomal 60S subunit protein involved in translation. YBR010W (\textit{HHT1}) is a histone protein required for chromatin assembly. YGL173C (\textit{XRN1}) is an exonuclease and deNADding enzyme that is involved in transcription initiation and elongation processes. YPL231W (\textit{FAS2}) is an alpha subunit of fatty acid synthetase involved in long-chain fatty acid biosynthesis. Finally, YER012W (\textit{PRE1}) is a beta 4 subunit of the 20S proteasome involved in the degradation of cellular proteins. From these proteins, YGL103W, YPL231W, YER012W are essential, while YDR497C, YGL256W, YBR010W, and YGL173C are non-essential (cf. SGD database and \cite{almeida_yeast_2021}). As such, these selected proteins play critical roles in diverse cellular processes, from central carbon metabolism to protein degradation, which explains their strong correlation with growth in our model.

\subsection{Specific modeling structure}
\label{subsec:model_structure_r}
As proof of concept for our proposed methodology, we considered the following model structure, which accounts only for growth and substrate uptake:
\begin{equation}
   \frac{\mathrm{d}{z_b}}{\mathrm{d}t} = \mu \cdot z_b, \, z_b(t_0) = z_{b_0},
\end{equation}
\begin{equation}
   \frac{\mathrm{d}{z_g}}{\mathrm{d}t} = q_g \cdot z_b, \, z_g(t_0) = z_{g_0},
\end{equation}
\begin{equation}
   \frac{\mathrm{d}{p_i}}{\mathrm{d}t} = 0, \, p_i(t_0) = p_{i_0}, \, \forall i \in \{1,2,...,n_{v_s} \},
\end{equation}
with the following kinetic functions:
\begin{equation}
    \mu = \frac{q_g - m_g}{\alpha},
    \label{eq:mu_function}
\end{equation}
\begin{equation}
    q_g = -q_{g,\mathrm{max}} \cdot \left( \frac{z_g}{z_g + k_g} \right).
\end{equation}
Here, the biomass and glucose concentrations are denoted by $z_b \in \mathrm{R}$ (in arbitrary units of $\mathrm{OD_{600}}$) and $z_g \in \mathbb{R}$ (in $\mathrm{g/L}$), respectively. The intracellular protein concentration of a selected feature $i$ is denoted by $p_i \in \mathbb{R}$. These protein concentrations are collected in the vector of selected features, $\bm{v_s}$. For simplicity, we assume that the selected proteins are at steady state, as the strains in the experiments are not subject to \textit{dynamic} perturbations of the proteome. Additionally, for the sake of demonstration, we use their initial (steady-state) concentrations directly from the original proteome dataset. We use Monod-type kinetics for the glucose uptake rate, $q_g$, following the derivation in \cite{heijnen_derivation_1995}, while the specific growth rate, $\mu$, is derived from Pirt’s equation for substrate distribution \cite{noauthor_maintenance_1965}. 

Regarding the model parameters, $q_{g,\text{max}}$ is the maximum biomass-specific substrate uptake, $k_g$ is a substrate affinity constant, $m_g$ is a maintenance-specific substrate uptake rate, and $\alpha$ a yield of substrate per unit biomass. We consider a subset of the model parameters as functions of the selected features, $\bm{\theta_f}(\bm{v_s}) := [q_{g,\text{max}}, \alpha, k_g]^\tran$, while $m_g$ is treated as a constant, $\bm{\theta_c} := m_g$. 

\begin{figure*}[htb!]
\centering
    \begin{subfigure}{0.32\textwidth}
      \captionsetup{justification=centering}
      \includegraphics[scale=0.6]{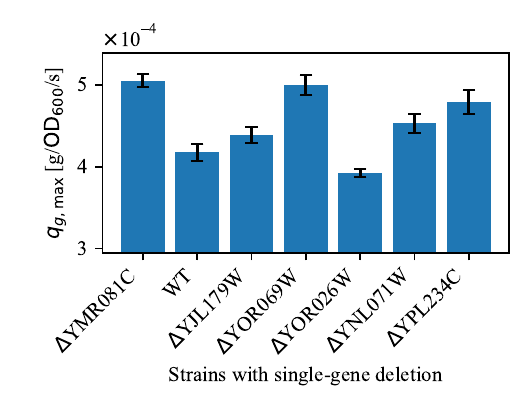}
      \subcaption[]{}
    \end{subfigure}
    \begin{subfigure}{0.32\textwidth}
      \captionsetup{justification=centering}
      \includegraphics[scale=0.6]{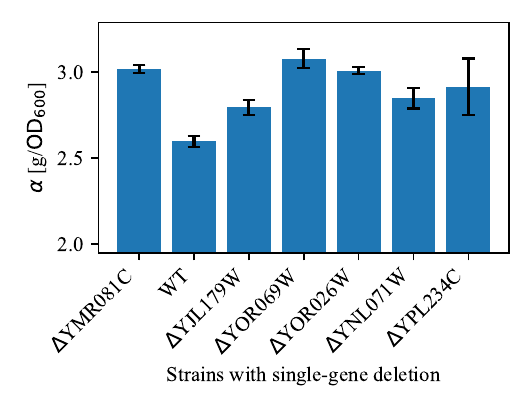}
      \subcaption[]{}
    \end{subfigure}
    \begin{subfigure}{0.32\textwidth}
      \captionsetup{justification=centering}
      \includegraphics[scale=0.6]{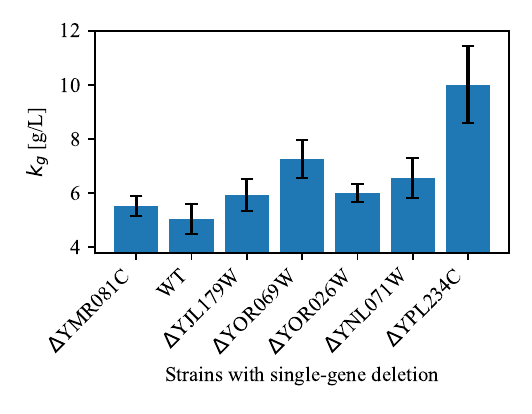}
      \subcaption[]{}
    \end{subfigure}
    
    \caption{Labels of the multi-output single-input Gaussian processes. Estimated parameter values of A) $q_{g,\text{max}}$, B) $\alpha$, and C) $k_g$ for the kinetic experiments involving the corresponding single-gene deletion strains. The deleted gene in the yeast strain is indicated by $\Delta$. The order of the strains, from left to right, is based on decreasing values of the reported growth rates (on agar medium) from the original database \cite{messner_proteomic_2023}.}
    \label{fig:labels_GPS}
\end{figure*}

Using Gaussian processes, the parameter values obtained from fitting each single-gene deletion experiment (Fig. \ref{fig:labels_GPS}) were mapped to the protein concentrations of the top seven most important features from the proteome database, identified using permutation importance (cf. Fig. \ref{fig:feature_importance_plot}). Note that these single-gene knockouts do not correspond to the genes of the selected features determined before. The single-gene deletion strains used in the kinetic experiments were specifically chosen as they portrayed well-distributed growth rates (on agar medium) across the original dataset's range, while the WT strain was used as a reference. We reasoned that these strains would thus provide a wide range of feature values for training the model.

\begin{figure*}[htb!]
\centering
    \begin{subfigure}{0.32\textwidth}
      \captionsetup{justification=centering}
      \includegraphics[scale=0.6]{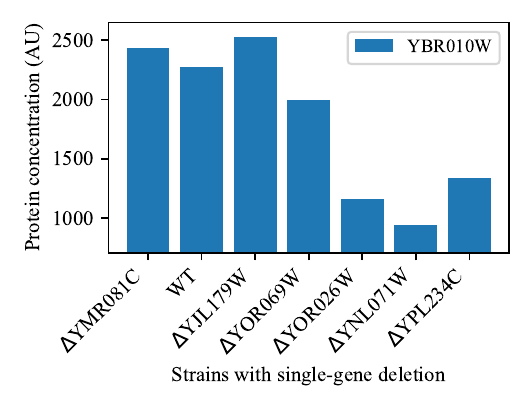}
      \subcaption[]{Protein feature YBR010W}
    \end{subfigure}
    \begin{subfigure}{0.32\textwidth}
      \captionsetup{justification=centering}
      \includegraphics[scale=0.6]{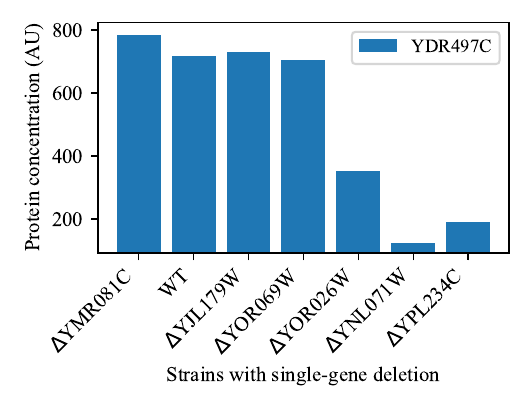}
      \subcaption[]{Protein feature YDR497C}
    \end{subfigure}
    \begin{subfigure}{0.32\textwidth}
      \captionsetup{justification=centering}
      \includegraphics[scale=0.6]{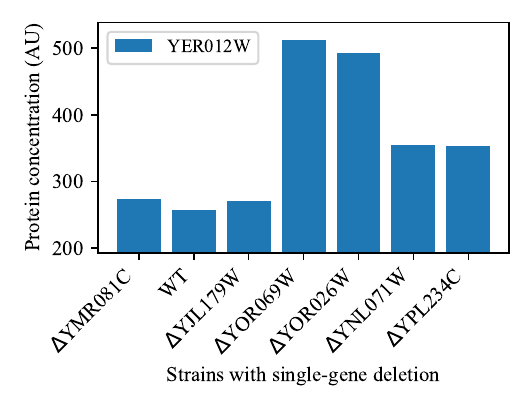}
      \subcaption[]{Protein feature YER012W}
    \end{subfigure}
    \\
    \begin{subfigure}{0.32\textwidth}
      \captionsetup{justification=centering}
      \includegraphics[scale=0.6]{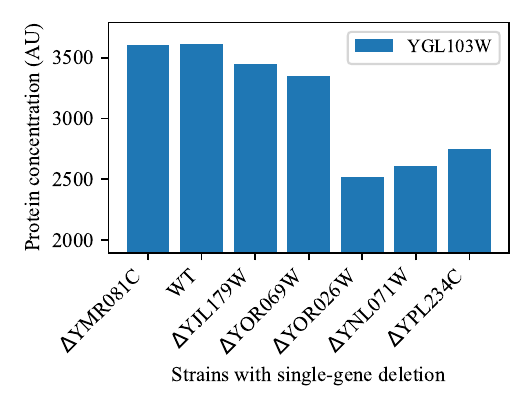}
      \subcaption[]{Protein feature YGL103W}
    \end{subfigure}
    \begin{subfigure}{0.32\textwidth}
      \captionsetup{justification=centering}
      \includegraphics[scale=0.6]{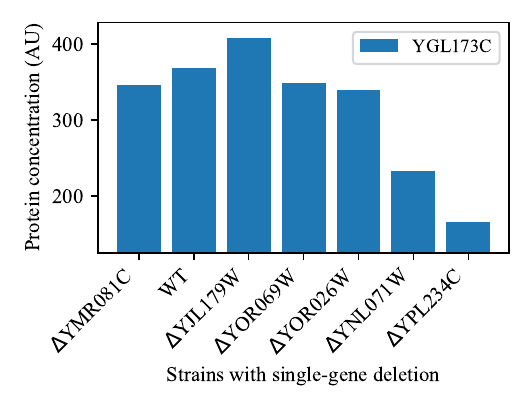}
      \subcaption[]{Protein feature YGL173C}
    \end{subfigure}
    \begin{subfigure}{0.32\textwidth}
      \captionsetup{justification=centering}
      \includegraphics[scale=0.6]{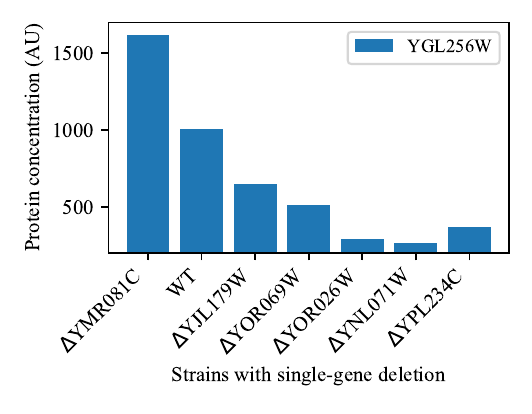}
      \subcaption[]{Protein feature YGL256W}
    \end{subfigure}
    \\
    \begin{subfigure}{0.32\textwidth}
      \captionsetup{justification=centering}
      \includegraphics[scale=0.6]{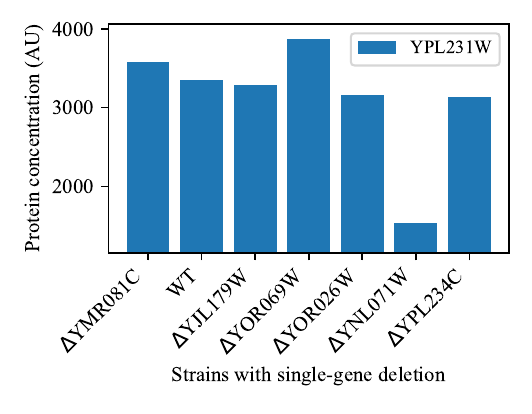}
      \subcaption[]{Protein feature YPL231W}
    \end{subfigure}   
    \caption{Features of the multi-output single-input Gaussian processes. Protein concentration values in arbitrary units of intensity for the kinetic experiments involving the corresponding single-gene deletion strains. The deleted gene in the yeast strain is indicated by $\Delta$. The order of the strains, from left to right, is based on decreasing values of the reported growth rates (on agar medium) from the original database \cite{messner_proteomic_2023}.}
    \label{fig:features_GPS}
\end{figure*}

When plotting the concentrations of these selected proteins (Fig. \ref{fig:features_GPS}), i.e., the features used by the Gaussian processes, one can observe indeed significant variation across the strains used in the kinetic experiments. Similarly, the parameter values also vary across strains, reflecting changes in the conditions tested, and further justifying the function-based nature of these parameters. The parameter $m_g = (5.3 \pm 1.3) \times 10^{-6} \, \mathrm{g/\mathrm{OD_{600}}/s}$  was obtained from fitting the WT strain and its mean value was considered constant across the models for the other strains. It is important to note that the parameter estimation was based on dynamic experimental data for biomass portraying exponential growth, while only the initial condition for glucose was used, as no dynamic concentrations for glucose was collected.

\subsubsection{Fitting of the hybrid model with Gaussian-process-predicted parameters}
\begin{figure*}[htb!]
    \centering
    \begin{subfigure}{0.4\textwidth}
      \captionsetup{justification=centering}
      \includegraphics[scale=0.5]{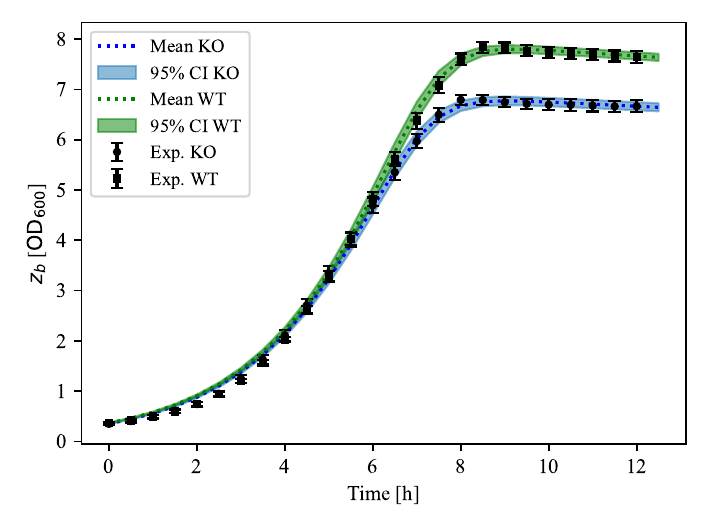}
      \subcaption[]{Biomass: $\Delta$YMR081C strain}
    \end{subfigure}
    \begin{subfigure}{0.4\textwidth}
      \captionsetup{justification=centering}
      \includegraphics[scale=0.5]{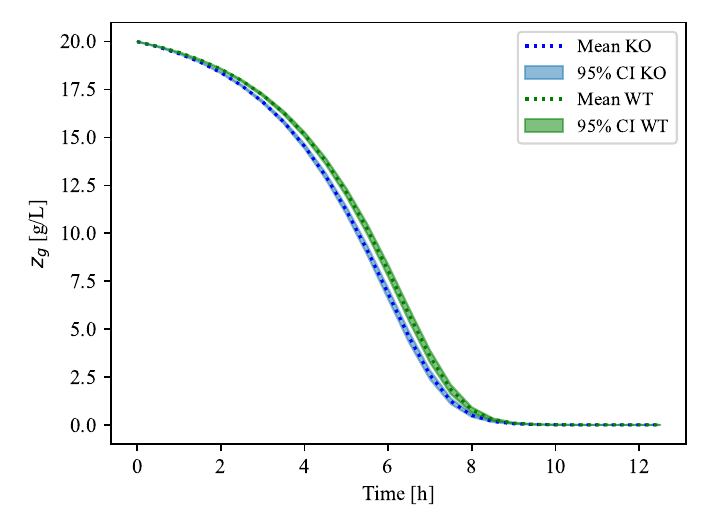}
      \subcaption[]{Glucose: $\Delta$YMR081C strain}
    \end{subfigure}

    \begin{subfigure}{0.4\textwidth}
      \captionsetup{justification=centering}
      \includegraphics[scale=0.5]{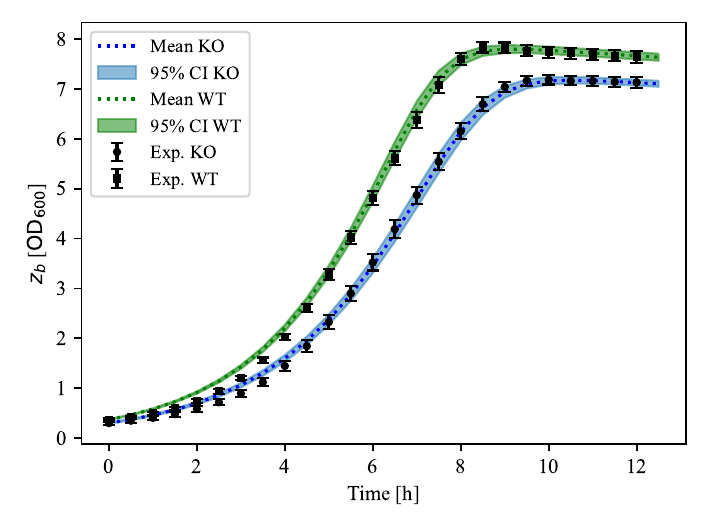}
      \subcaption[]{Biomass: $\Delta$YJL179W strain}
    \end{subfigure}
    \begin{subfigure}{0.4\textwidth}
      \captionsetup{justification=centering}
      \includegraphics[scale=0.5]{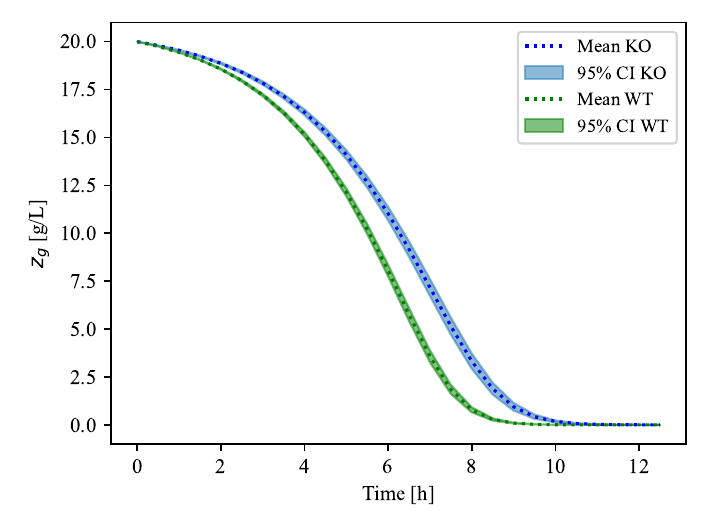}
      \subcaption[]{Glucose: $\Delta$YJL179W strain}
    \end{subfigure}

    \begin{subfigure}{0.4\textwidth}
      \captionsetup{justification=centering}
      \includegraphics[scale=0.5]{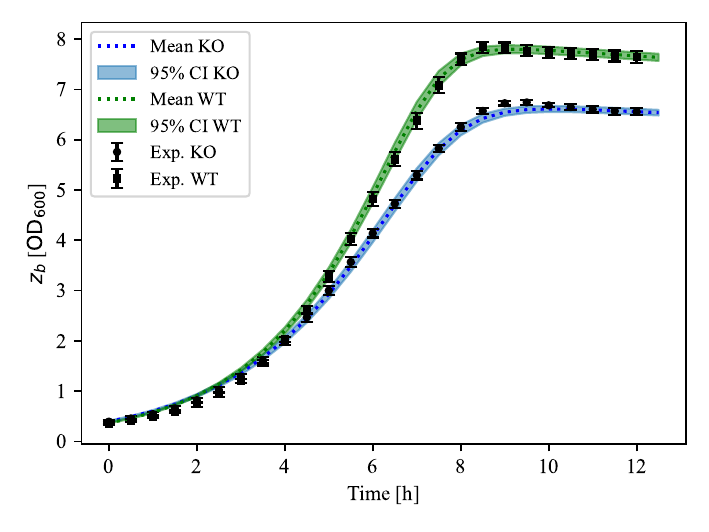}
      \subcaption[]{Biomass: $\Delta$YOR069W strain}
    \end{subfigure}
    \begin{subfigure}{0.4\textwidth}
      \captionsetup{justification=centering}
      \includegraphics[scale=0.5]{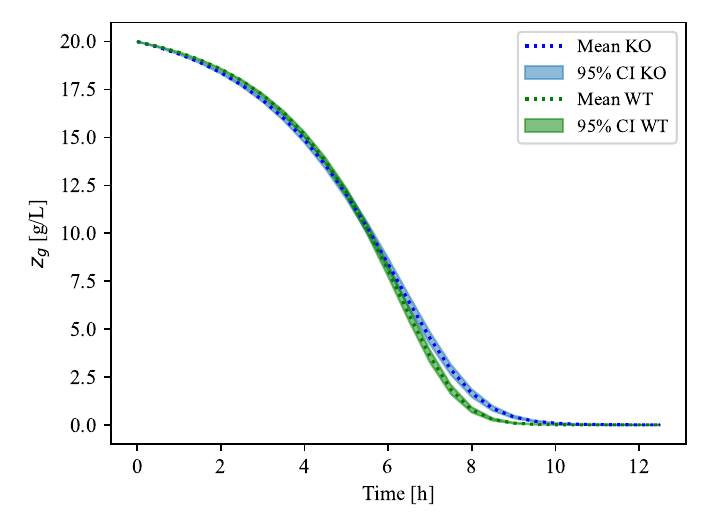}
      \subcaption[]{Glucose: $\Delta$YOR069W strain}
    \end{subfigure}
    \caption{Fitting of the dynamic experiments with the hybrid model using Gaussian-process-predicted parameters. Experimental data (Exp.) is shown for both the single-gene deletion strains ($\Delta$) and the wild-type (WT). The dotted lines represent mean predictions, with the 95 \% confidence interval (CI) indicated by the shaded areas.}
    \label{fig:dyn_exp_1}
\end{figure*}

\begin{figure*}[htb!]
    \centering
    \begin{subfigure}{0.4\textwidth}
      \captionsetup{justification=centering}
      \includegraphics[scale=0.5]{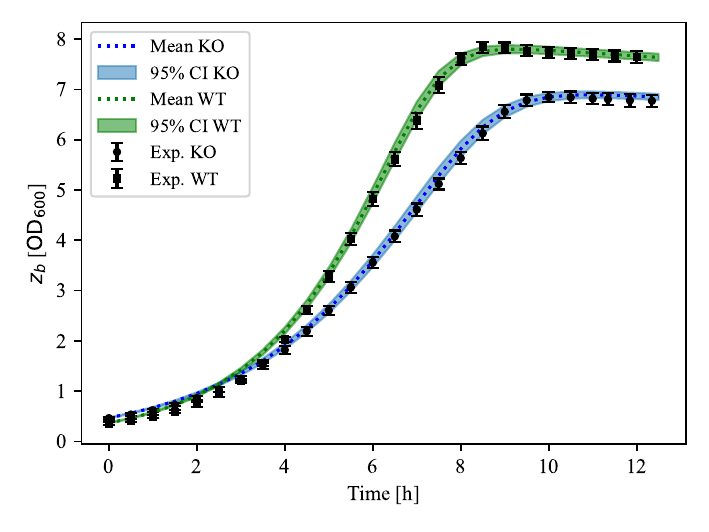}
      \subcaption[]{Biomass: $\Delta$YOR026W strain}
    \end{subfigure}
    \begin{subfigure}{0.4\textwidth}
      \captionsetup{justification=centering}
      \includegraphics[scale=0.5]{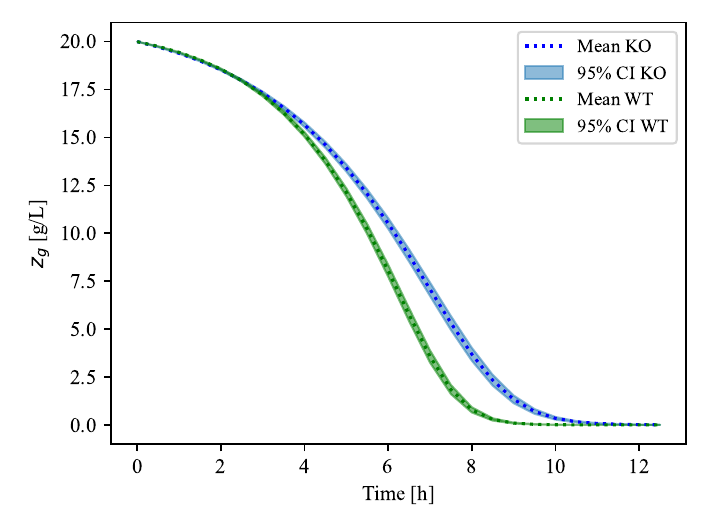}
      \subcaption[]{Glucose: $\Delta$YOR026W strain}
    \end{subfigure}

    \begin{subfigure}{0.4\textwidth}
      \captionsetup{justification=centering}
      \includegraphics[scale=0.5]{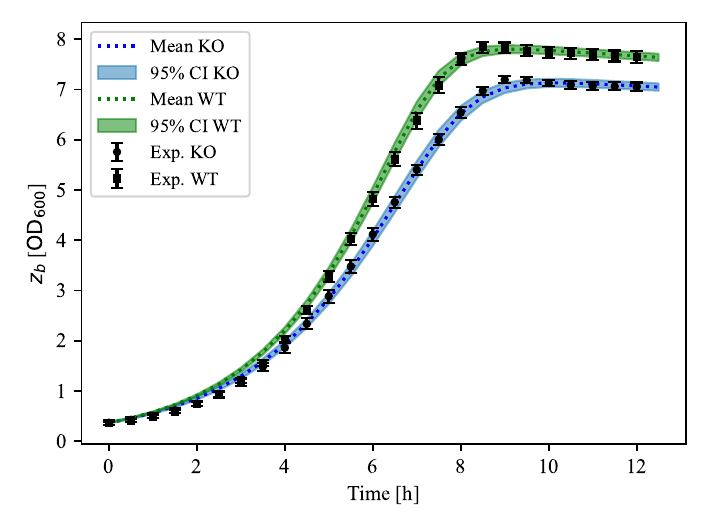}
      \subcaption[]{Biomass: $\Delta$YNL071W strain}
    \end{subfigure}
    \begin{subfigure}{0.4\textwidth}
      \captionsetup{justification=centering}
      \includegraphics[scale=0.5]{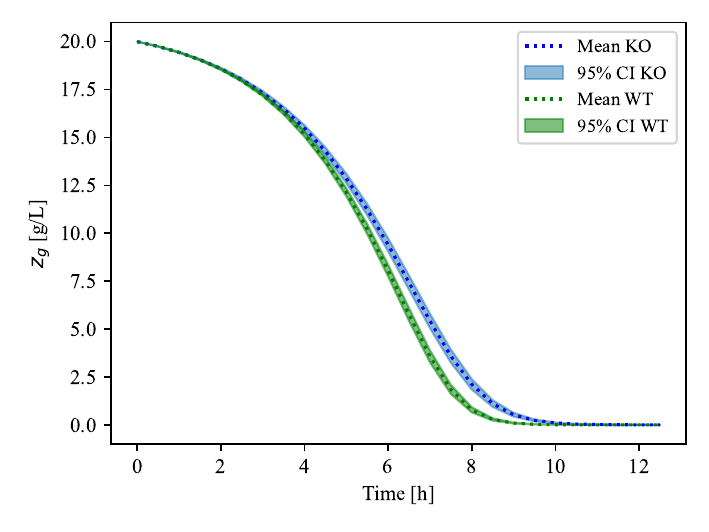}
      \subcaption[]{Glucose: $\Delta$YNL071W strain}
    \end{subfigure}

    \begin{subfigure}{0.4\textwidth}
      \captionsetup{justification=centering}
      \includegraphics[scale=0.5]{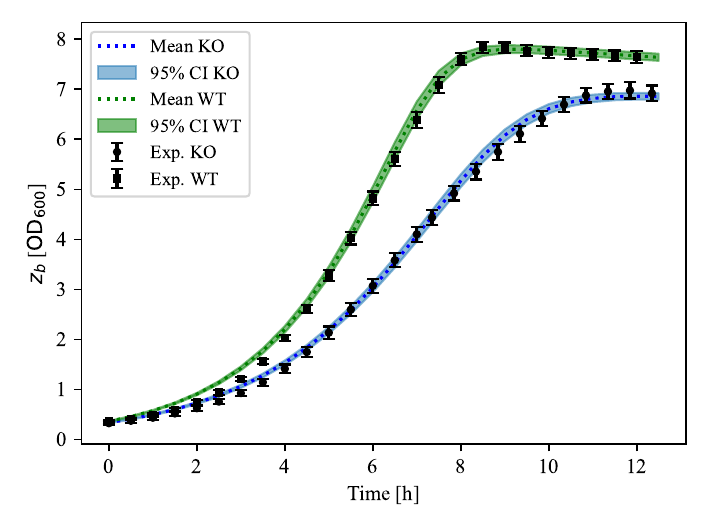}
      \subcaption[]{Biomass: $\Delta$YPL234C strain}
    \end{subfigure}
    \begin{subfigure}{0.4\textwidth}
      \captionsetup{justification=centering}
      \includegraphics[scale=0.5]{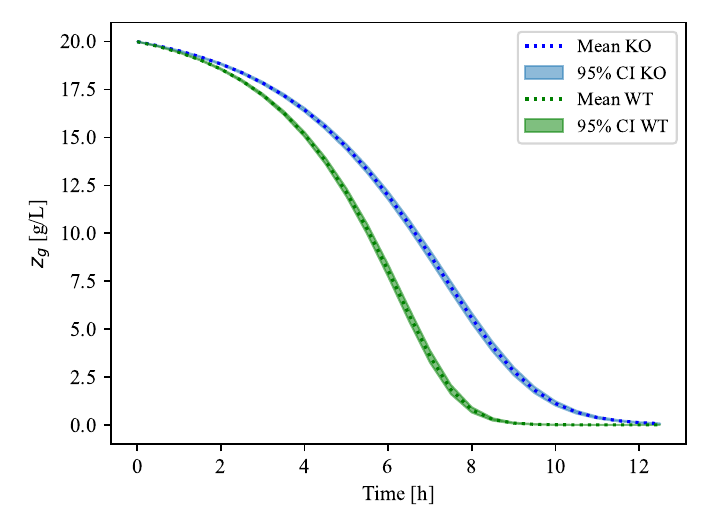}
      \subcaption[]{Glucose: $\Delta$YPL234C strain}
    \end{subfigure}

    \caption{Fitting of the dynamic experiments with the hybrid model using Gaussian-process-predicted parameters (\textbf{continued}). Experimental data (Exp.) is shown for both the single-gene deletion strains ($\Delta$) and the wild-type (WT). The dotted lines represent mean predictions, with the 95 \% confidence interval (CI) indicated by the shaded areas.}
    \label{fig:dyn_exp_2}
\end{figure*}

The dynamic experiments, along with the fitting of the hybrid model after training the Gaussian processes to predict the parameters in $\bm{\theta_f}$, are presented in Figs. \ref{fig:dyn_exp_1}-\ref{fig:dyn_exp_2}. Overall, the hybrid model fit the experimental data with satisfactory performance, with the results falling within the estimated 95 \% confidence interval of the model. The predicted glucose concentration profiles, although not measured, show the expected hyperbolic behavior dictated by Monod-type kinetics. The maintenance coefficient in Eq. \eqref{eq:mu_function} captures the slight decrease in biomass concentration after glucose depletion. {Furthermore, $\alpha$ helps to capture differences in biomass-to-substrate yields (i.e., $1/\alpha$) that result from changes in metabolic flux distributions associated with varying proteome profiles.}

\begin{figure*}[htb!]
\centering
  \includegraphics[scale=0.68]{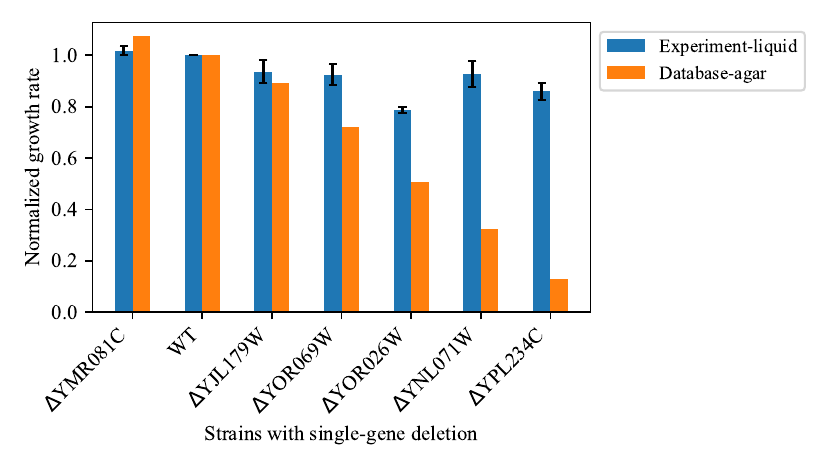}
   \caption{Normalized maximum specific growth rate, relative to the WT strain, for dynamic experiments in liquid medium, compared with values from the database on agar medium. The order of the strains, from left to right, is based on decreasing values of the reported growth rates (on agar medium) from the original database \cite{messner_proteomic_2023}.}
  \label{fig:experiments_metrics}
\end{figure*}

When comparing the normalized growth rates (relative to the WT strain) for the kinetic experiments in liquid medium with those from the original database (agar medium), there is a similar overall decreasing trend following the gene deletions of YMR081C, YJL179W, YOR069W, and YOR026W (Fig. \ref{fig:experiments_metrics}).  However, this trend is not followed by the gene deletions of YNL071W and YPL234C, although their growth rates remain lower than the {WT-normalized growth. As expected, the growth rates are generally higher in liquid media due to better mass transfer and substrate availability. This discrepancy in growth across experimental setups (agar vs. liquid medium), along with the varying biomass yields not captured by the proteomics/agar-based growth database, justifies our decision not to use the database \textit{directly} for building the hybrid dynamic model. Nevertheless, the latter database still provides a well-grounded starting point or initial guess for feature selection and dimensionality reduction through the conducted analysis with random forests, which can facilitate multiscale modeling in systems biology. In fact, in many cases, previously published omics datasets are likely to be generated under conditions different from those of the intended applications, but it would still be desirable to mine and extract value from those datasets.}

{By linking the parameters to proteomic features through Gaussian processes, and upon training with \textit{more} data, our approach has the potential to enable generalization to unseen data, predicting dynamic growth for different proteome profiles. Although the parameter values are derived from strain-specific proteomic data, the selected proteomic features correlate with the observed phenotypes, as demonstrated by our analysis with random forests. This indicates that the underlying biological processes influencing growth in \textit{S. cerevisiae} can be effectively captured through changes in these features.}

\begin{figure*}[htb!]
\centering
    \begin{subfigure}{0.32\textwidth}
      \captionsetup{justification=centering}
      \includegraphics[scale=0.6]{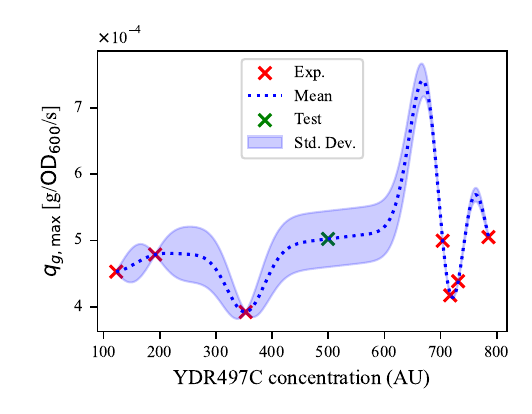}
      \subcaption[]{}
    \end{subfigure}
    \begin{subfigure}{0.32\textwidth}
      \captionsetup{justification=centering}
      \includegraphics[scale=0.6]{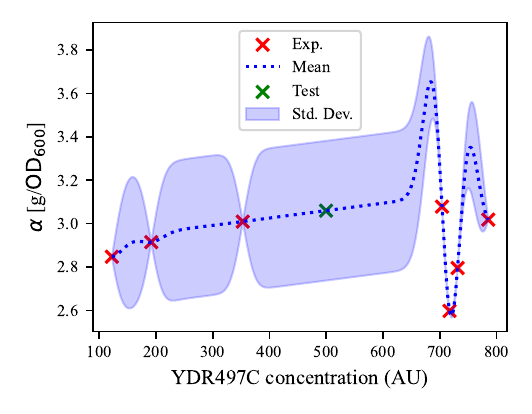}
      \subcaption[]{}
    \end{subfigure}
    \begin{subfigure}{0.32\textwidth}
      \captionsetup{justification=centering}
      \includegraphics[scale=0.6]{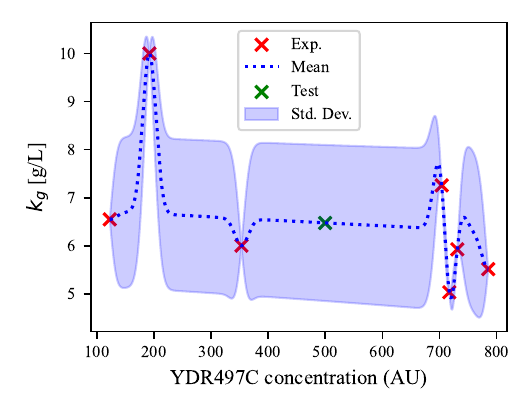}
      \subcaption[]{}
    \end{subfigure}
    \begin{subfigure}{0.4\textwidth}
      \captionsetup{justification=centering}
      \includegraphics[scale=0.5]{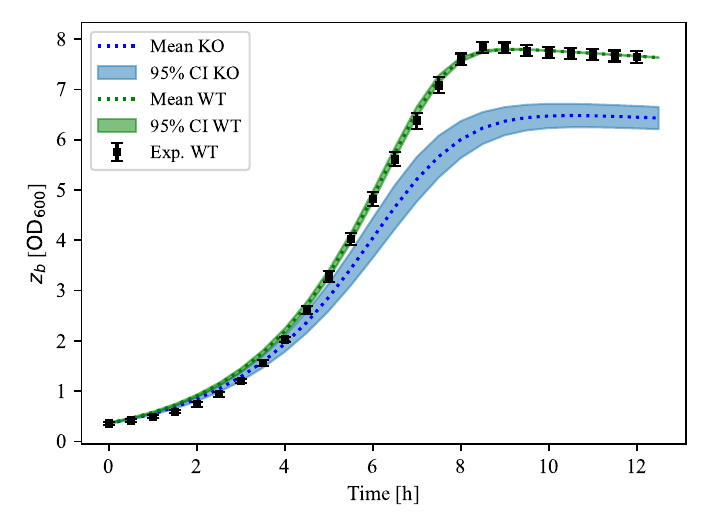}
      \subcaption[]{Biomass: Hypothetical strain}
    \end{subfigure}
    \begin{subfigure}{0.4\textwidth}
      \captionsetup{justification=centering}
      \includegraphics[scale=0.5]{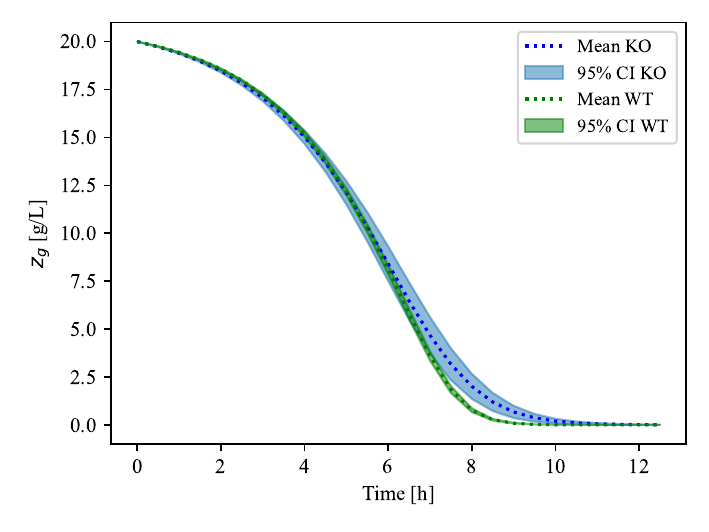}
      \subcaption[]{Glucose: Hypothetical strain}
    \end{subfigure}
    \caption{{Fitting of a single-input single-output Gaussian process with YDR497C concentration as the input feature (the top important feature, cf. Fig. \ref{fig:feature_importance_plot}) and the parameters of the dynamic model as outputs labels: A) $q_{g,\text{max}}$, B) $\alpha$, and C) $k_g$. The predicted mean and standard deviation (Std. Dev.) are shown. A test feature value for a hypothetical strain is also indicated. Here, we considered a \textit{linear prior mean} $\mathfrak{m}(\bm{v_s})=\beta_1{\theta_f}_i(\bm{v_s}) + \beta_2$, where $\beta_1$ and $\beta_2$ are the coefficient and bias of the linear function, respectively. D)-E) Prediction of the hybrid model for the hypothetical strain using the single-feature Gaussian-process-predicted parameters. Experimental data (Exp.) is shown only for the wild-type strain (WT). The dotted lines represent mean predictions, with the 95 \% confidence interval (CI) indicated by the shaded areas.}}
    \label{fig:GPS_single_feat}
\end{figure*}

As previously discussed, the {current} model has limitations, primarily due to the small dataset used for training the parameter functions through the Gaussian processes. Specifically, only seven sets of parameters were available for training, corresponding to the number of kinetic experiments, and since each Gaussian-process-predicted parameter depends on seven features (protein concentrations), the limited dataset with respect to the feature space significantly restricts the model's ability to confidently predict beyond the experimental data. {In other words, the Gaussian processes may exhibit overfitting due to the limited training dataset. As shown in Fig. \ref{fig:GPS_single_feat} A-C, even a single feature, e.g., YDR497C concentration (the top important feature in Fig. \ref{fig:feature_importance_plot}), is sufficient to fit the dataset \textit{as it is now}. As more data becomes available, particularly in high-throughput settings, additional features will likely be essential to capture correlations to the labels that a single feature cannot, as supported by our random forest analysis in Fig. \ref{fig:feat_coeff_det_bar_plot}.}

{The high confidence in the predictions of the dynamic model based on the seven-feature Gaussian processes (cf. Figs. \ref{fig:dyn_exp_1}-\ref{fig:dyn_exp_2}) arises because the currently available parameters are sampled from regions of the feature space where data is available. This high confidence is visually apparent in the single-feature Gaussian processes in Fig. \ref{fig:GPS_single_feat} A-C, where uncertainty is low at the training data points. Naturally, the predicted uncertainty may increase when the model encounters unseen feature values or if future kinetic experiments reveal varying label values for the same feature values. As a proof of concept, to illustrate a scenario with higher predicted uncertainty, we use the single-feature Gaussian processes from Fig. \ref{fig:GPS_single_feat} A-C to build the hybrid model, assuming an arbitrary YDR497C concentration of 500 AU for a hypothetical strain with no prior experimental data. We maintain the same initial-condition uncertainty as in the wild-type (WT) strain experiment. The resulting dynamic behavior is shown in Fig. \ref{fig:GPS_single_feat} D-E. We focus on the single-feature Gaussian processes here for simplicity of representation (two dimensions) and to exemplify how this uncertainty propagates over time. As expected, the confidence interval for the dynamic trajectory in this hypothetical scenario is larger than, for example, that of the WT strain, where prior data is available to train the Gaussian processes.}

As Gaussian processes are data-driven methods, they will naturally perform better with more experimental data, such as in high-throughput settings. Therefore, the presented hybrid model can be further enhanced by incorporating more data. A key advantage of our approach is that future experiments can be more targeted. For example, by actively exploring areas of high uncertainty, batch-to-batch experiments can be designed to improve the model iteratively. The reduced dimensionality of the selected omics features simplifies the size of the explorable feature space, making the approach more efficient. 

With more experimental data, one may {also} find that it is necessary to include additional features to improve prediction accuracy, or conversely, that one can reduce the feature space even further. The prediction accuracy of the Gaussian process regressors could be employed during this second round of feature selection to guide this decision. Furthermore, while we used Gaussian processes as the machine-learning part of the hybrid model due to the small dataset size and its ability to easily estimate uncertainty, the training of Gaussian processes can become computationally expensive as the dataset grows, since the computational cost scales cubically with the number of experiments, $\mathcal{O}(n_e^3)$ \cite{rasmussen_gaussian_2006}. In such cases, alternative machine learning methods, such as deep neural networks, may be more suitable for handling larger datasets. For neural networks, uncertainty estimation can be performed leveraging, for example, maximum likelihood, approximate Bayesian, or bootstrapping methods \cite{papadopoulos_confidence_2001}.

\section{Conclusion}
In this work, we presented a hybrid dynamic modeling framework that integrates various machine-learning tools to facilitate feature selection and dimensionality reduction in omics datasets, aiming to build continuous and differentiable data-driven functions that link intracellular processes to the resulting phenotype. As proof of concept, we applied random forests to a previously published high-dimensional proteomics dataset, identifying key intracellular proteins via permutation feature importance that correlate with \textit{S. cerevisiae} growth. With a reduced feature space, targeted experiments for dynamic modeling can be conducted. Upon performing these experiments, key parameters were modeled as functions of the selected protein concentrations using Gaussian processes, aiming to capture the dynamic behavior of yeast strains under different proteome profiles. The resulting hybrid model, incorporating Gaussian-process-predicted parameters, provided accurate predictions of the dynamic trajectories with uncertainty estimation.

Despite the limited size of the dataset for the dynamic experiments, our results demonstrate the potential of this pipeline to mine omics data for feature selection in multiscale biological models. The data-driven nature of the hybrid modeling strategy promises to enhance predictive power and generalizability as more experimental data is available. {This includes the option to refine the model, for instance by adjusting the number of features to balance predictive accuracy and overfitting; yet the current approach already provides a strong starting point}. While Gaussian processes were selected for their uncertainty estimation capabilities, alternative machine-learning approaches, such as deep neural networks, could be explored with larger datasets, particularly in high-throughput settings. Overall, this study illustrates how integrating omics data with hybrid modeling has the potential to efficiently inform multiscale modeling approaches in systems biology and bioprocess engineering. {Future work will involve cases with more layers of omics data, e.g., fluxomics, to account for formation of product and other metabolites, potentially unlocking advanced \textit{model-based} biotechnological processes and applications.}
\newline

\noindent \textbf{CRediT authorship contribution statement} \newline
\textbf{Sebastián Espinel-Ríos}: Conceptualization, Methodology, Software, Formal Analysis, Writing - Original Draft, Writing - Review \& Editing, Visualization.
\textbf{José Montaño López}: Investigation, Validation,  Writing - Original Draft, Writing - Review \& Editing.
\textbf{José L. Avalos}: Conceptualization, Writing - Original Draft, Writing - Review \& Editing, Supervision, Funding Acquisition. \newline

\noindent \textbf{Declaration of interests} \newline
The authors declare that they have no conflict of interest. \newline

\noindent \textbf{Data statement} \newline
Data will be made available on request. \newline

\noindent \textbf{Acknowledgment} \newline
This work was supported by the U.S. DOE-BER Grant DESC0022155 and the U.S. NSF Grant MCB-2300239.\newline

\bibliographystyle{elsarticle-num} 
\bibliography{bibliography}

\end{document}